# Non-Standard Neutrino Interactions in the mu-tau sector

Irina Mocioiu[*]

*Department of Physics, The Pennsylvania State University, University Park, PA 16802, USA*

*Kavli Institute for Theoretical Physics, University of California, Santa Barbara, California 93106, USA*

Warren Wright[†]

*Department of Physics, The Pennsylvania State University, University Park, PA 16802, USA*

(Dated: October 21, 2014)

We discuss the effects of non-standard neutrino interactions on muon rates in high statistics atmospheric neutrino oscillation experiments like IceCube DeepCore. We concentrate on the mu-tau sector, which is presently the least constrained. It is shown that the magnitude of the effects depends strongly on the sign of the $\epsilon_{\mu\tau}$ parameter describing this non-standard interactions. A simple analytic model is used to understand the parameter space where differences between the two signs are maximized. We discuss how this effect is partially degenerate with changing the neutrino mass hierarchy, as well as how this degeneracy could be lifted.

## I.   INTRODUCTION

One of the major goals of the present and future generation of neutrino experiments is to firmly establish the correct framework for our understanding of neutrinos and their interactions.

While in the last decades we have learned a lot about neutrino mass differences and mixings, their origin remains poorly understood. An extension to the Standard Model is required to accommodate such neutrino properties and in many instances it leads to additional, non-standard neutrino interactions. Neutrino oscillations can be affected by the addition of Non-Standard Interactions (NSI) and can thus be used to look for physics beyond the standard model. NSI have been considered from the first discussion

———
[*]Electronic address: irina@phys.psu.edu
[†]Electronic address: wright@psu.edu

of matter effects in [1] and many subsequent analysis [2]. Present data already constrains NSI [3, 4], but a large parameter space is still unexplored. NSI might even explain some sub-leading oscillation phenomena [5].

Studies of the future long baseline neutrino experiments have shown that they can achieve very high precision measurements, at the few percent level [6]. Under the assumption of three neutrino flavor mixing and standard model interactions, this translates into precision measurements of the neutrino parameters, which can be a useful guide towards model building. In the presence of additional neutrino species or non-standard interactions, this might however no longer be the case, due to the large degeneracies between standard and non-standard oscillation parameters. Consistency checks between different oscillation measurements thus become extremely valuable. While fixed baseline beam experiments can provide very high precision, atmospheric neutrinos probe a much larger range of possible standard and non-standard neutrino properties parameter space as they cover a much broader energy range and can take advantage of large matter effects when traversing the Earth core. Many of the systematic uncertainties associated with the production and detection of atmospheric neutrinos can be overcome by the extremely high number of events expected in future very large neutrino detectors, by using some spectral and directional information.

The IceCube DeepCore (ICDC) project [7] is a low energy extension of IceCube that has been developed in the last few years. This was originally motivated by searches for neutrinos of astrophysical origin, but would collect huge numbers of atmospheric neutrinos that can be used for a series of interesting physics measurements of neutrino oscillations: extracting the mass hierarchy [8], detecting tau neutrinos [9] and better measurements of main neutrino oscillation parameters [10]. Studies of atmospheric neutrinos and their oscillations have now become an important objective for ICDC [11] and the collaboration is considering an extension of the low energy reach of IceCube DeepCore with the Precision IceCube Next Generation Upgrade (PINGU) [12]. In addition to extending sensitivity at lower energies, PINGU will improve sensitivity at higher energies as well, by allowing for better reconstruction and characterization of events. Some studies of these and other future experiments' sensitivity to non-standard interactions have been explored in [13].

In this paper we explore the effects of non-standard neutrino interactions on atmospheric neutrino oscillations observed in large detectors like the IceCube DeepCore. We concentrate on the mu-tau sector which is the least constrained. In addition ICDC might offer the first opportunity for detecting a large sample of tau neutrinos [9], though we will not directly use this information in our analysis. We primarily explore energies above 10GeV where ICDC already has good sensitivity that can be further improved should PINGU be built. This energy range allows for much better directional reconstruction of the muons and is in the deep inelastic interaction region, where cross-sections are well understood and additional kinematic information might be available.

In Sec. II we will review the framework used for analyzing effects of non-standard interactions on neutrino oscillations, as well as the constraints present data imposes on standard and NSI parameters. In Sec. III we will present the results of numerical simulations showing the effect of the NSI parameter $\epsilon_{\mu\tau}$ on muon neutrino survival probability and number of events observable in large atmospheric neutrino detectors like IceCube DeepCore. These results show a sign asymmetry which we describe analytically in Sec. IV. In Sec.V we discuss the degeneracy between the sign of $\epsilon_{\mu\tau}$ and the neutrino mass hierarchy, as well as how this degeneracy could be addressed. The conclusions are presented in Sec. VI.

## II. THEORETICAL FRAMEWORK

Over the last decade many different types of experiments have provided evidence for neutrino oscillations and have allowed mapping the allowed regions for three flavor neutrino oscillation parameters. These are two mass square differences, $\Delta m_{21}^2$, $\Delta m_{31}^2$, three mixing angles, $\theta_{12}$, $\theta_{23}$, $\theta_{13}$, and a CP-violating phase $\delta$. The mixing matrix connecting the flavor basis ($\nu_{e,\mu,\tau}$) and the mass basis ($\nu_{1,2,3}$) can be written in terms of these parameters as:

$$U = \begin{pmatrix} 1 & 0 & 0 \\ 0 & c_{23} & s_{23} \\ 0 & -s_{23} & c_{23} \end{pmatrix} \begin{pmatrix} c_{13} & 0 & s_{13}e^{-i\delta_{cp}} \\ 0 & 1 & 0 \\ -s_{13}e^{i\delta_{cp}} & 0 & c_{13} \end{pmatrix} \begin{pmatrix} c_{12} & s_{12} & 0 \\ -s_{12} & c_{12} & 0 \\ 0 & 0 & 1 \end{pmatrix},$$

with $c_{ij} = \cos\theta_{ij}$ and $s_{ij} = \sin\theta_{ij}$.

Neutrino propagation is affected by their interactions with the matter they traverse. Standard model neutral current interactions are the same for the three active neutrinos so they do not affect neutrino oscillations. Charged-current interactions of the electron type neutrinos on electrons will however generate an effective matter potential:

$$V_{cc} = \sqrt{2}G_F N_e = 7.6 \times 10^{14} Y_e \rho,$$

where $G_F$ is the Fermi constant, $N_e$ is the electron number density, $Y_e$ is the electron fraction and $\rho$ is the density of the medium neutrinos propagate through. For many experiments the propagation medium is the Earth and the Earth's density can be approximated using the PREM model [14].

Since the origin of any potential non-standard neutrino interactions is still unknown, we consider a general parametrization in terms of $\epsilon_{\alpha\beta}$ where $\alpha$ and $\beta$ indicate neutrino flavor. The matter potential in the presence of NSI is parametrized as:

$$V_{eff} = \begin{pmatrix} 1 + \epsilon_{ee} & |\epsilon_{e\mu}|e^{i\delta_{e\mu}} & |\epsilon_{e\tau}|e^{i\delta_{e\tau}} \\ |\epsilon_{e\mu}|e^{-i\delta_{e\mu}} & \epsilon_{\mu\mu} & |\epsilon_{\mu\tau}|e^{i\delta_{\mu\tau}} \\ |\epsilon_{e\tau}|e^{-i\delta_{e\tau}} & |\epsilon_{\mu\tau}|e^{-i\delta_{\mu\tau}} & \epsilon_{\tau\tau} \end{pmatrix} \tag{1}$$

The NSI contributions can arise from exchange of new heavy particles or through more complicated interactions with hidden sectors. They are parametrized by an effective Lagrangian:

$$\mathcal{L}_{NSI} = -2\sqrt{2}G_F\bar{\nu}_\alpha\gamma_\mu\nu_\beta \left(\epsilon_L^{\alpha\beta,ij}\bar{f}_L^i\gamma^\mu f_L^j + \epsilon_R^{\alpha\beta,ij}\bar{f}_R^i\gamma^\mu f_R^j\right) \tag{2}$$

where $\epsilon_{L,R}^{\alpha\beta,ij}$ describe the coupling of neutrinos of flavors $\alpha$ and $\beta$ to the left (right)-handed components of fermions $f_i$ and $f_j$. When summing over all possible fermions $f_i$ and $f_j$ present in matter (e.g. electrons, up and down quarks inside the Earth), the parameters $\epsilon_{\alpha\beta}$ in eq. (1) are obtained. For any given new physics model it should be possible to map the model parameters that determine $\epsilon_{L,R}^{\alpha\beta,ij}$ onto $\epsilon_{\alpha\beta}$ which can be constrained by data.

Non-standard neutrino interactions can affect neutrino oscillations in multiple ways: in the production, detection and propagation of neutrinos. In this paper we consider propagation of atmospheric neutrinos that can be detected in IceCube DeepCore or the possible future PINGU detector.

Most NSI parameters have been constrained by experimental observations, together with measurements of the standard oscillation parameters. Best-fit values for the mixing matrix parameters [15] and current model independent bounds for the NSI parameters [3] are given in Table I. The bounds for the NSI parameters assume neutral Earth-like matter with an equal number of neutrons and protons.

| Mass and Mixing | Best Fit $\pm 1\sigma$ | NSI Bounds |
|---|---|---|
| $\sin^2(\theta_{12})$ | $0.302^{+0.013}_{-0.012}$ | $|\epsilon_{ee}| \leq 4.2$ |
| $\sin^2(\theta_{23})$ | $0.413^{+0.037}_{-0.025} \oplus 0.594^{+0.021}_{-0.022}$ | $|\epsilon_{e\mu}| \leq 0.33$ |
| $\sin^2(\theta_{13})$ | $0.0227^{+0.0023}_{-0.0024}$ | $|\epsilon_{e\tau}| \leq 3.0$ |
| $\Delta m^2_{21}$ | $7.50^{+0.18}_{-0.19} \times 10^{-5}eV^2$ | $|\epsilon_{\mu\mu}| \leq 0.068$ |
| $\Delta m^2_{31}(N)$ | $+2.473^{+0.070}_{-0.067} \times 10^{-3}eV^2$ | $|\epsilon_{\mu\tau}| \leq 0.33$ |
| $\Delta m^2_{32}(I)$ | $-2.427^{+0.042}_{-0.065} \times 10^{-3}eV^2$ | $|\epsilon_{\tau\tau}| \leq 21$ |

TABLE I: Mixing matrix best fit values and NSI parameters bounds.

For an experiment like IceCube's DeepCore (ICDC), the total number of muons, $N_\mu$, can be obtained from:

$$N_\mu\left(\Delta E_\mu, \Delta\theta\right) = 2\pi t N_A \int_{E_{\mu,i}}^{E_{\mu,f}} dE_\mu \int_{\theta_i}^{\theta_f} \sin\theta d\theta \int_{E_\mu}^{\infty} dE_\nu M\left(E_\nu\right) \frac{\partial\sigma_{\nu_\mu}^{CC}}{\partial E_\nu}\left(E_\nu, E_\mu\right) \times$$
$$\left(\frac{\partial^2\phi_{\nu_\mu}\left(E_\nu,\theta\right)}{\partial E_\nu\partial\theta} P_{\mu\mu}\left(E_\nu,\theta\right) + \frac{\partial^2\phi_{\nu_e}\left(E_\nu,\theta\right)}{\partial E_\nu\partial\theta} P_{e\mu}\left(E_\nu,\theta\right)\right). \tag{3}$$

$E_{\mu,i}$ and $E_{\mu,f}$ define the observed muon energy bin, $\theta_i$ and $\theta_f$ define the zenith angle bin, $t$ is the time period during which the detector was taking data, $M$ is the effective mass of the detector, and $N_A$ is Avogadro's number. The cross section[16] for neutrino interaction in the detector is given by $\sigma_{\nu_\mu}^{CC}$ and the flux[17] of neutrinos is given by $\phi_\nu$. The effective mass of ICDC is taken from [7] and displayed in Figure 1.

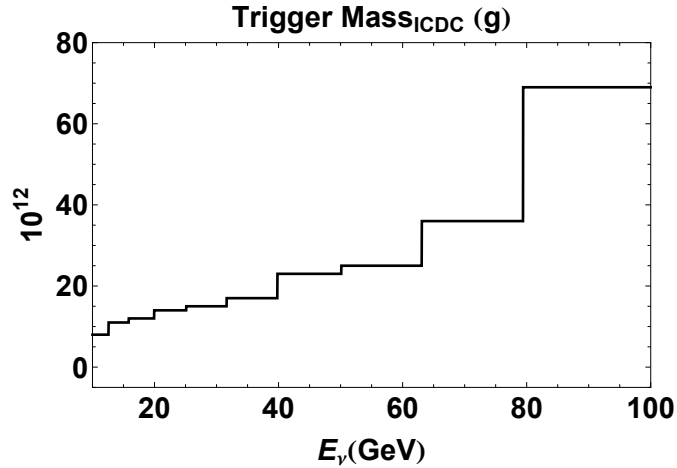

FIG. 1: Effective Mass of ICDC.

## III.  NUMERICAL RESULTS

We first study the effects of $\epsilon_{\mu\tau}$ on oscillation probabilites. To first approximation we set all phases to zero and all $\epsilon = 0$ except for $\epsilon_{\mu\tau}$. In Figure 2 the effects of small changes in $\epsilon_{\mu\tau}$ can be seen on the oscillation probabilities for normal and inverted mass hierarchies. For even small deviations of $\epsilon_{\mu\tau}$ from $\epsilon_{\mu\tau} = 0$, the changes to $P_{\mu\mu}$ can be large. One remarkable feature of Figure 2 is that the sign of $\epsilon_{\mu\tau}$ has a significant effect on $P_{\mu\mu}$. In fact, it seems, at least qualitatively, that flipping the sign of $\epsilon_{\mu\tau}$ is nearly equivalent to changing the mass hierarchy. This is an effect that will be examined in section V.

The effects of $\epsilon_{\mu\tau}$ can be further seen when calculating the number of events via Equation 3. $N_\mu$ has numerically been calculated for energy bins 5 GeV wide and for the three zenith angle bins given in Table II. These cover the Earth core, with the highest densities and density variations and two more large angular bins. It is likely the experiment can achieve better angular reconstruction in the long term, but this is probably a realistic starting point and it accounts for all the smearing due to both detector resolution and the spread between observed muon and incoming neutrino direction.

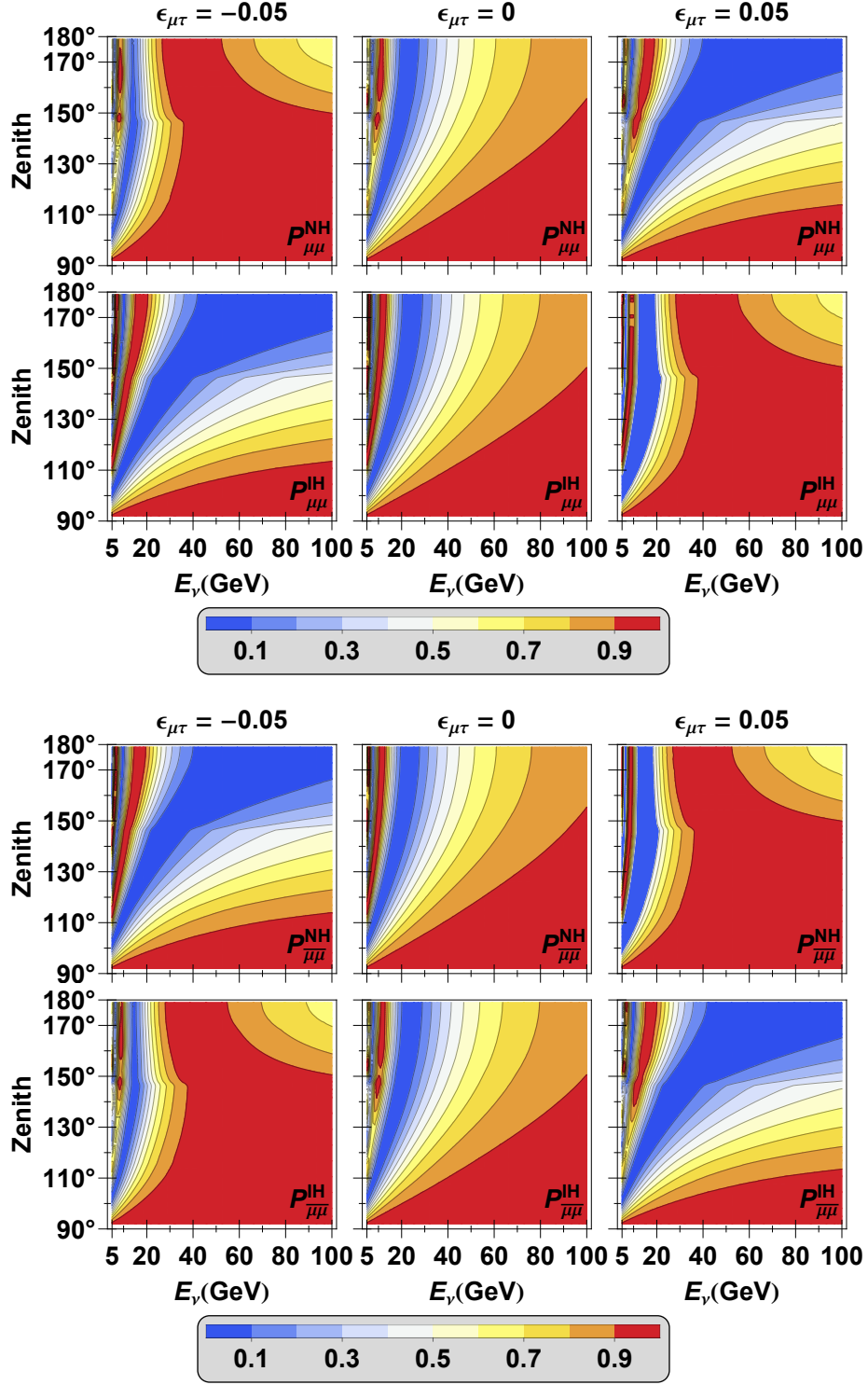

FIG. 2: $\epsilon_{\mu\tau}$ effects on $P_{\mu\mu}$.

Figures 3 and 4 show the effect $\epsilon_{\mu\tau}$ has on the total number of muon neutrino events in the detector.

| | Core ($\Delta\theta_1$) | Mantle ($\Delta\theta_2$) | Outer Mantle/Crust ($\Delta\theta_3$) |
|---|---|---|---|
| $\cos(\theta)$ Range: | $(-1, -0.837)$ | $(-0.837, -0.446)$ | $(-0.446, 0)$ |

TABLE II: Zenith bins.

It is clear that $\epsilon_{\mu\tau}$ has a much greater effect on $N_\mu$ for muons that are produced along trajectories that go through the dense core. Furthermore, it is evident that the large effect of the sign of $\epsilon_{\mu\tau}$ is also strongest for neutrino trajectories traversing the core. This sign dependence is also visible in the mantle and crust but the effect is less dramatic.

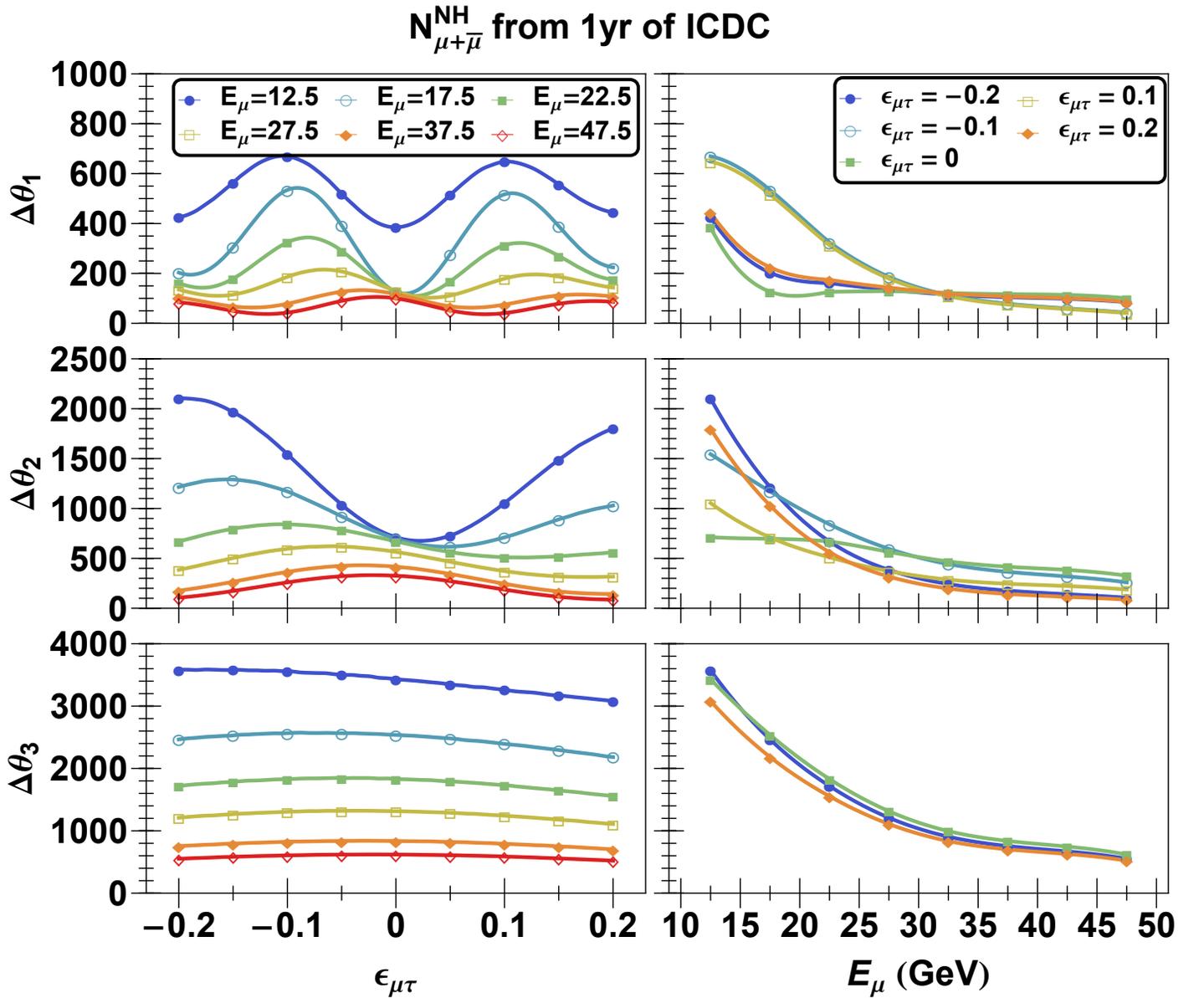

FIG. 3: $\epsilon_{\mu\tau}$ effects on the number of muons for normal mass hierarchy.

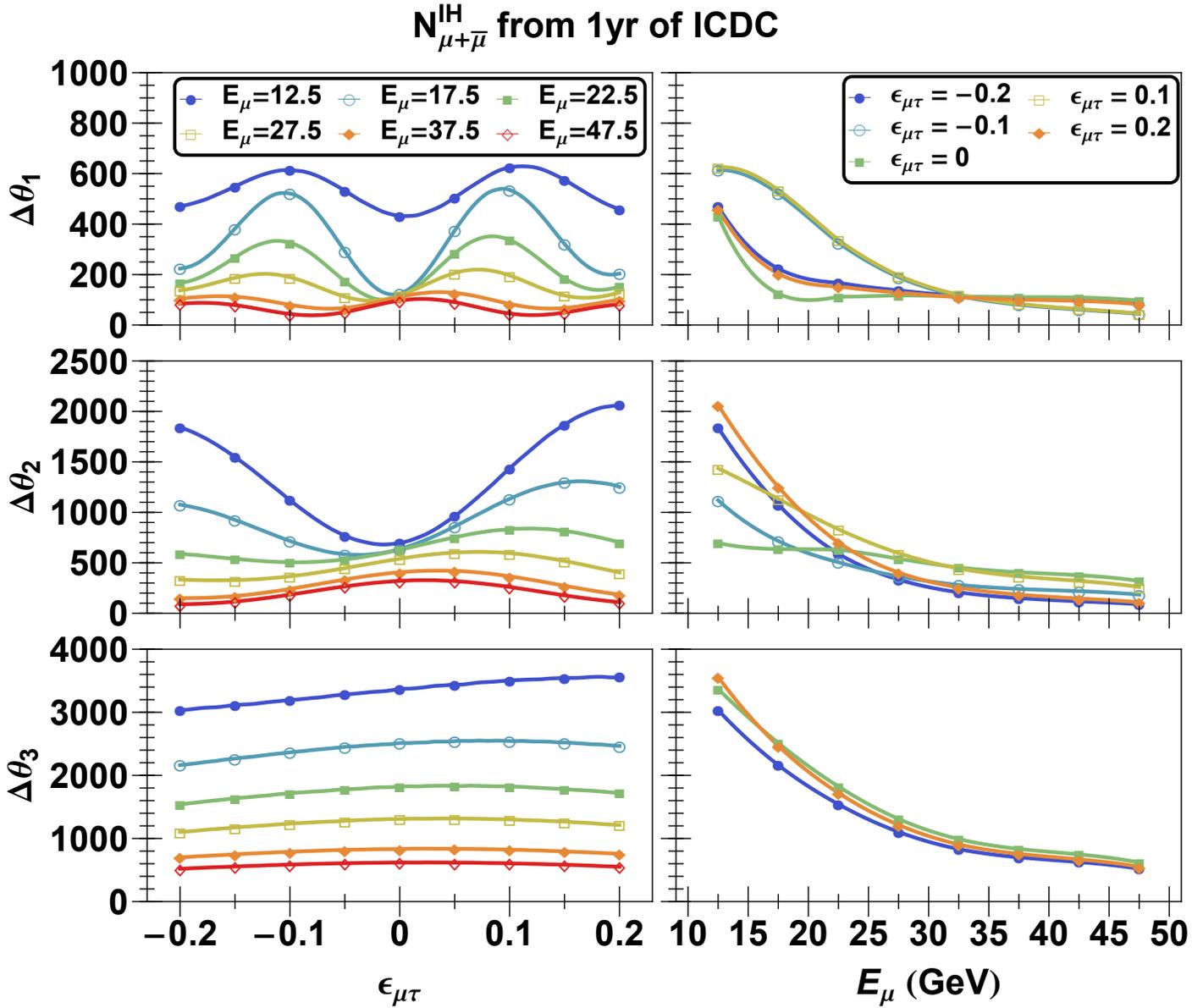

FIG. 4: $\epsilon_{\mu\tau}$ effects on the number of muons for inverted mass hierarchy.

## IV. $\epsilon_{\mu\tau}$ ANALYSIS: REGIONS OF MAXIMAL SIGN DEPENDENCE

As seen in Figures 2, 3, and 4, the effects of $\epsilon_{\mu\tau}$ on $P_{\mu\mu}$ and $N_\mu$ demonstrate a sign dependence. This sign dependence is in contrast to the other NSI parameters where the sign of the parameter does not have nearly as significant an effect. In order to analytically analyze this dependence we set all $\delta = 0$ and all $\epsilon = 0$ except $\epsilon_{\mu\tau}$. Furthermore, we set

$$\Delta m_{21}^2 = \theta_{12} = \theta_{13} = \delta_{cp} = 0 \text{ and } \theta_{23} = \pi/4,$$

and we assume a constant density in $V_{cc}$. It should be noted that these are assumptions made only in our analytical approximations that help us better track and understand the effects of the sign of $\epsilon_{\mu\tau}$ and thus any value gained from them would be qualitative only. Our numerical results are based on the full realistic parameter values and density profiles and they show that, below 15 GeV the analytical model's usefulness drops rather dramatically. These assumptions lead to a simplified Schrödinger-like equation:

$$
i\frac{d}{dx}\begin{pmatrix} \nu_e(x) \\ \nu_\mu(x) \\ \nu_\tau(x) \end{pmatrix} = \begin{pmatrix} V_{cc} - \frac{\Delta m_{31}^2}{4E_\nu} & 0 & 0 \\ 0 & 0 & \frac{\Delta m_{31}^2}{4E_\nu} + V_{cc}\epsilon_{\mu\tau} \\ 0 & \frac{\Delta m_{31}^2}{4E_\nu} + V_{cc}\epsilon_{\mu\tau} & 0 \end{pmatrix}\begin{pmatrix} \nu_e(x) \\ \nu_\mu(x) \\ \nu_\tau(x) \end{pmatrix}
$$

This equation can then be solved to yield:

$$
\begin{pmatrix} \nu_e(L) \\ \nu_\mu(L) \\ \nu_\tau(L) \end{pmatrix} = \begin{pmatrix} e^{-\frac{1}{4}iL\left(4V_{cc}-\frac{\Delta m_{31}^2}{E_\nu}\right)} & 0 & 0 \\ 0 & \cos(L\Lambda) & -i\sin(L\Lambda) \\ 0 & -i\sin(L\Lambda) & \cos(L\Lambda) \end{pmatrix}\begin{pmatrix} \nu_e(0) \\ \nu_\mu(0) \\ \nu_\tau(0) \end{pmatrix}
$$

with $\Lambda = \frac{\Delta m_{31}^2}{4E_\nu} + V_{cc}\epsilon_{\mu\tau}$. Thus:

$$
P_{\mu\mu} = \cos^2\left(L\left(\frac{\Delta m_{31}^2}{4E_\nu} + V_{cc}\epsilon_{\mu\tau}\right)\right) \tag{4}
$$

A useful measure of the sign symmetry is:

$$
\Delta_\epsilon P_{\mu\mu} = P_{\mu\mu}(\epsilon_{\mu\tau}) - P_{\mu\mu}(-\epsilon_{\mu\tau}) = -\sin\left(2L\frac{\Delta m_{31}^2}{4E_\nu}\right)\sin(2LV_{cc}\epsilon_{\mu\tau})
$$

Thus $|\Delta_\epsilon P_{\mu\mu}|$ assumes maximum values when:

$$
2L\frac{\Delta m_{31}^2}{4E_\nu} = (2n+1)\frac{\pi}{2}
$$

$$
2LV_{cc}\epsilon_{\mu\tau} = (2m+1)\frac{\pi}{2}
$$

where $m,n \in \mathbb{Z}$ and $m,n \geq 0$.

From the above analysis, equations can be derived that specify the location of maximum sign asymmetry:

$$
\therefore E_\nu = \left(\frac{2m+1}{2n+1}\right)\frac{\Delta m_{31}^2}{4V_{cc}\epsilon_{\mu\tau}} \tag{5}
$$

$$
\therefore L = \frac{(2m+1)\pi}{4V_{cc}\epsilon_{\mu\tau}} \tag{6}
$$

$$
\text{also } L = \frac{(2n+1)\pi}{\Delta m_{31}^2}E_\nu \tag{7}
$$

where:

$$
L = 2R_{Earth}\sin\left(\theta - \frac{\pi}{2}\right) \tag{8}
$$

By using Equation 8, Equations 5 and 6 will predict the neutrino energy and direction where the sign of $\epsilon_{\mu\tau}$ has the most effect on muon neutrino survival probabilities and detected number of muon neutrino events, and Equation 7 defines curves that pass through these points.

The predictions of Equations 5, 6, and 7, are displayed in Figure 5. Figure 5 shows $\Delta_\epsilon P_{\mu\mu}$ for two different values of $\epsilon_{\mu\tau}$. Equation 7's predictions are represented by the two white on black lines and the white on black stars represent the predictions of Equations 5 and 6. We see that these predications coincide remarkably well with the maxima and minima of a numerically calculated $\Delta_\epsilon P_{\mu\mu}$.

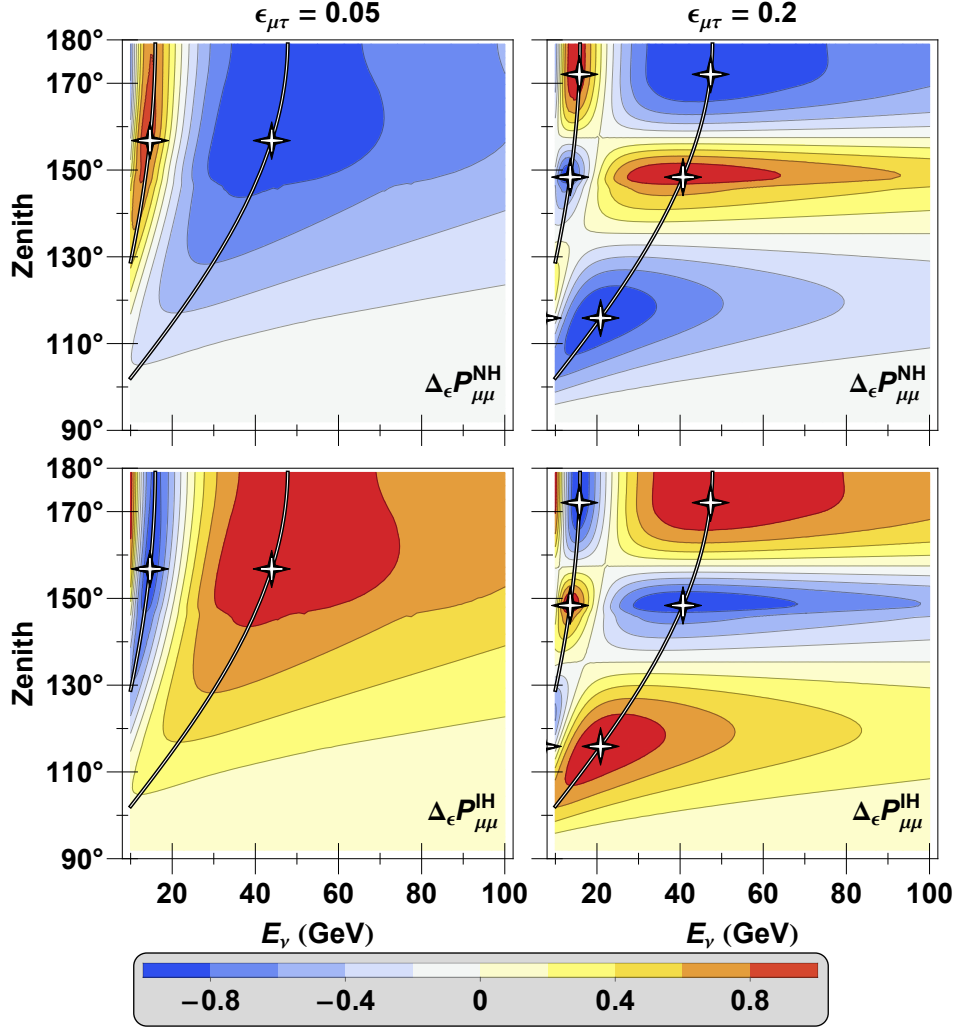

FIG. 5: NSI asymmetry prediction

The significance of the predictive power of this analytical model is twofold. Firstly, it demonstrates that the sign dependent features observed in $P_{\mu\mu}$ and $N_\mu$ are not numerical artifacts and that they can be qualitatively described by a simple model. Secondly, the predictions themselves are useful in defining parameter space regions where the analysis of experimental results would best determine the sign of $\epsilon_{\mu\tau}$.

## V. MASS HERARCHY IMPLICATIONS

As noted in Section III, the effect of changing the sign of $\epsilon_{\mu\tau}$ seems similar to the effect of changing the mass hierarchy. This similarity becomes obvious upon the examination of Equation 4. It shows that, for the model under consideration, changing the sign of $\epsilon_{\mu\tau}$ is mathematically equivalent to changing the mass hierarchy. This is exactly true in the reduced model discussed in section IV, but it is only approximately true in a realistic model. One of the consequences of this feature is the uncertainty encountered in the fitting of a particular mass hierarchy model to experimental results. For instance, an experimentally obtained $N_\mu\left(E_\mu\right)$ curve might look like either $N_\mu^{NH}\left(\epsilon_{\mu\tau}=0\right)$ or $N_\mu^{IH}\left(\epsilon_{\mu\tau}\neq 0\right)$. Figure 6 shows how $N_\mu^{NH}\left(\epsilon_{\mu\tau}=0\right)$ and $N_\mu^{IH}\left(\epsilon_{\mu\tau}=0.00743\right)$ look very similar and how $N_\mu^{NH}\left(\epsilon_{\mu\tau}=0.00555\right)$ or $N_\mu^{IH}\left(\epsilon_{\mu\tau}=0\right)$ look very similar. This mass hierarchy - sign of NSI degeneracy effect shows how even a very small NSI value can make the determination of mass hierarchy even more difficult.

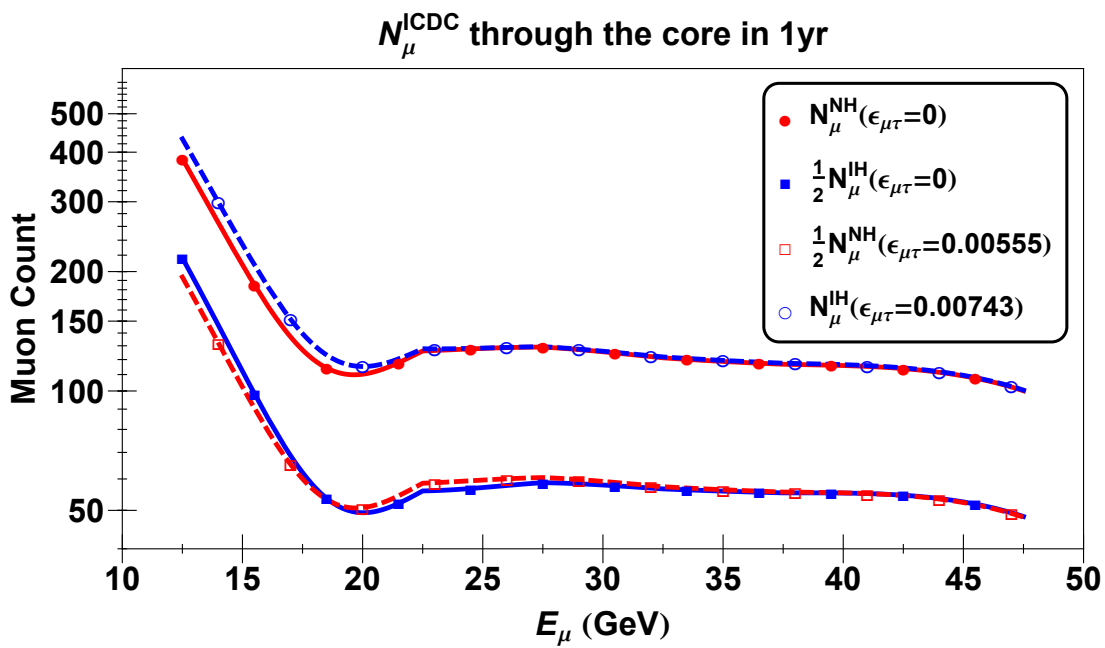

FIG. 6: Mass hierarchy imitation by specific NSI values.

This degeneracy is further illustrated in Figure 7. The plot compares a 10% measurement of the *normal hierarchy* under the assumption that standard three flavor oscillations are entirely responsible for the signal with the expected signal for an *inverted hierarchy* in the presence of a very small negative non-standard interaction parameter. The $N_\mu^{IH}\left(E_\mu, 0<\epsilon_{\mu\tau}<0.01\right)$ curve seems to be within 10% of the $N_\mu^{NH}\left(E_\mu, \epsilon_{\mu\tau}=0\right)$ curve. Both Figures 6 and 7 show the total number of muons and antimuons that would be detected at ICDC in one year of running for muons traversing the Earth's core.

The feature that Figures 6 and 7 illustrate is the degeneracy between mass hierarchy and the sign of $\epsilon_{\mu\tau}$. It is also useful to note that Figures 6 and 7 reveal that this degeneracy seems more prominent at higher energies. Below 20 GeV, the degeneracy is less pronounced, but above 20 GeV the match becomes very clear. This degeneracy has serious implications when attempting to fit theoretical models to experimentally determined spectra. Consider the task of deciding which mass hierarchy best describes the nature of neutrino oscillations. If $\epsilon_{\mu\tau} = 0$ is assumed, then a certain level of experimental precision is needed to distinguish a NH muon spectrum from an IH muon spectrum. But if $\epsilon_{\mu\tau} = 0$ is not assumed, then a much greater level of experimental precision is needed because of the degeneracy between mass hierarchy and the sign of $\epsilon_{\mu\tau}$. The degree of precision needed can be quantitatively determined by performing a $\chi^2$ analysis.

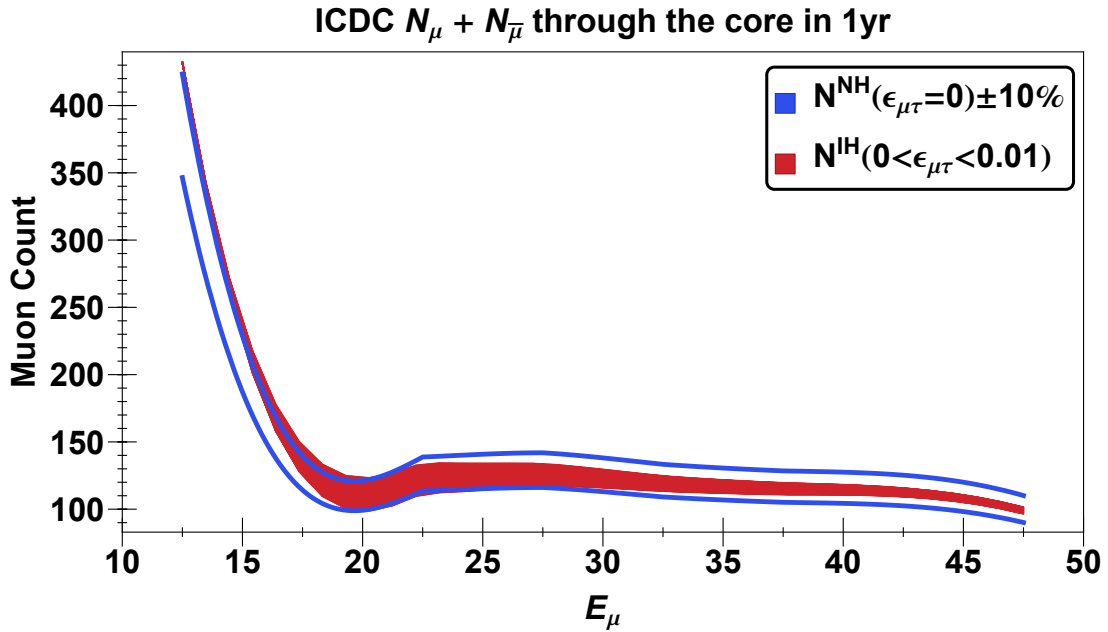

FIG. 7: Mass hierarchy imitation by a range of NSI values.

The test statistic under consideration is:

$$\chi^2\left(\epsilon_{\mu\tau}\right) = \sum_{E_\mu} \frac{N_{\mu+\bar{\mu}}^{\text{Model}}\left(E_\mu, \epsilon_{\mu\tau}\right) - N_{\mu+\bar{\mu}}^{\text{Null}}\left(E_\mu, \epsilon_{\mu\tau} = 0\right)}{N_{\mu+\bar{\mu}}^{\text{Null}}\left(E_\mu, \epsilon_{\mu\tau} = 0\right)} \tag{9}$$

Where Model and Null denote a mass hierarchy choice, either NH or IH. This test statistic, together with energy and angular binning, will represent a quantitative sensitivity or discovery potential analysis. In order to demonstrate how the degeneracy between the mass hierarchy and the sign of $\epsilon_{\mu\tau}$ can be broken, the muon energy spectrum will be divided into a low energy bin and a high energy bin. The two angular bins, Core and Mantle will be dealt with separately.

Figures 8 and 9 show the results of the $\chi^2$ analysis for muons traversing the core. The left plots represent the one year, low energy analysis and the right plots show the five year, high energy analysis. On each plot the horizontal lines represent different confidence levels and the different curves represent different Model and Null assignments as per the legend. Figure 8 analyzes the case of distinguishing a true NH three-flavor scenario from the effects of NSIs. Figure 9 analyzes the case of distinguishing IH.

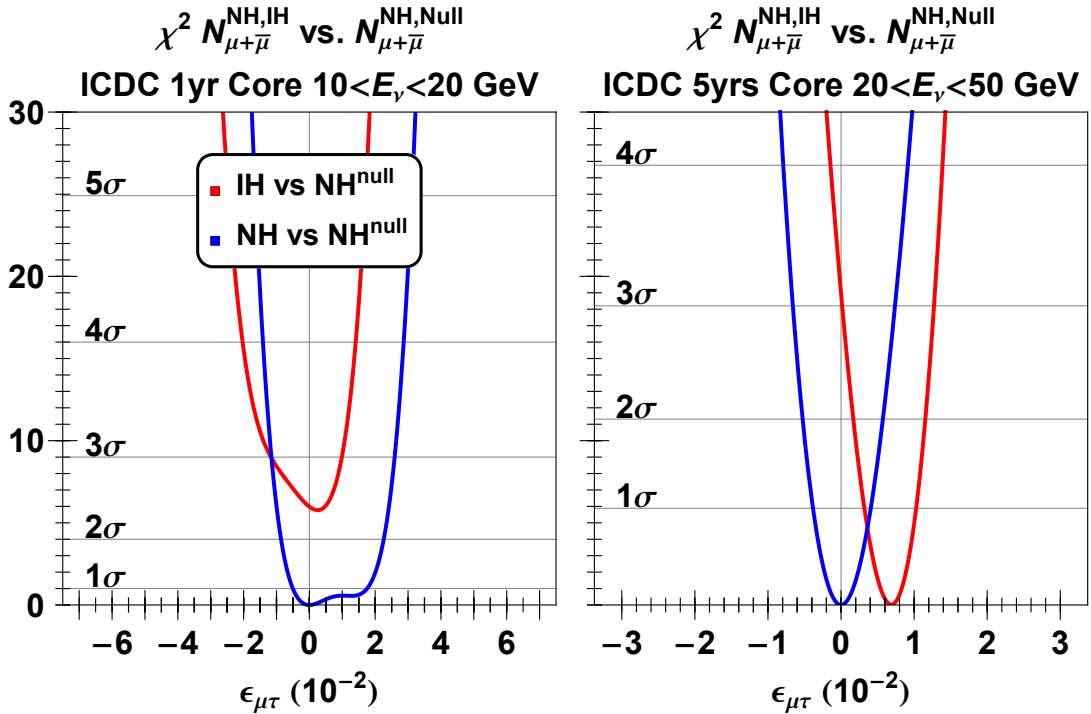

FIG. 8: $\chi^2$ analysis for NH true and for muons traversing the Core.

Consider the right graph in Figure 8. As is expected, the NH vs. NH$^{\text{Null}}$ curve shows that even at $1\sigma$ there is still a range of $\epsilon_{\mu\tau}$ which give rise to a muon spectrum that is indistinguishable from the $\epsilon_{\mu\tau} = 0$ case. Similarly, if NH and $\epsilon_{\mu\tau} = 0$ is true, then IH and $\epsilon_{\mu\tau} = 0$ can be ruled out at approximately a $3\sigma$ confidence level from the high energy analysis and it can be ruled out at approximately a $2.5\sigma$ confidence level from the low energy analysis. But, if NH and $\epsilon_{\mu\tau} = 0$ is true, then while IH and $\epsilon_{\mu\tau} \neq 0$ can be ruled out at approximately a $2.5\sigma$ confidence level from the low energy analysis, it cannot be ruled out at any confidence level from the high energy analysis. This shows, as expected, that the degeneracy between the mass hierarchy and the sign of $\epsilon_{\mu\tau}$ can be disentangled by only considering the low energy bins.

The situation is qualitatively unchanged if IH and $\epsilon_{\mu\tau} = 0$ is true. The effects of the degeneracy are even more important in this case, as the matter effect enhancements are now present for antineutrinos which are the sub-dominant component in the total number of events. Figure 9 shows that, even for the low energy analysis, NH and $\epsilon_{\mu\tau} = 0$ can be ruled out at approximately $2.5\sigma$ but NH and $\epsilon_{\mu\tau} \neq 0$ can only be ruled out at approximately $1.3\sigma$ confidence level. For the high energy analysis, the slight raising of the NH vs. IH$^{\text{Null}}$ curve compared to the IH vs. NH$^{\text{Null}}$ curve from Figure 8 gives meager hope that a high energy sample could break the degeneracy, given enough time.

Figures 10 and 11 show the results of the $\chi^2$ analysis for muons traversing the mantle. The structure

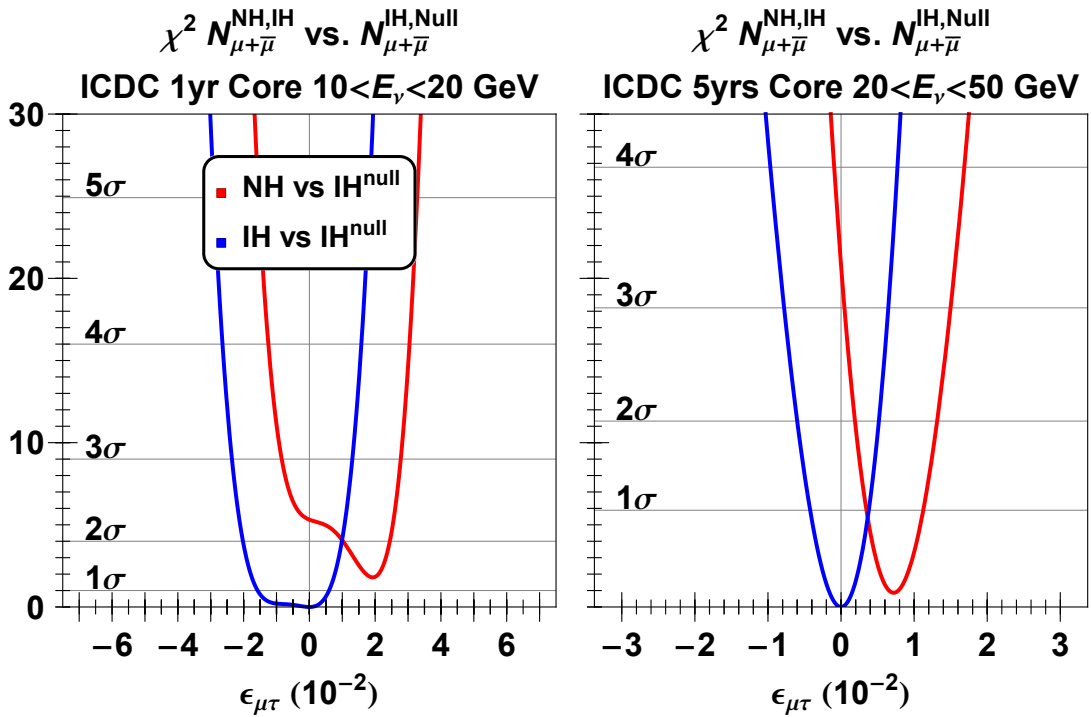

FIG. 9: $\chi^2$ analysis for IH true and for muons traversing the Core.

and layout of these figures is the same as for Figures 8 and 9.

The only significant difference between the plots of Figures 10 and 11 from Figures 8 and 9, is that the sensitivity curves are widened because of the decrease in the matter effect coming from the mantle compared with the core. This widening represents a decrease in the predictive power of the analysis and thus the core analysis is preferred.

Figures 8 - 11 show that the degeneracy between mass hierarchy and the sign of $\epsilon_{\mu\tau}$ can be broken by focusing on muons that traverse the Earth's core and have an energy below 20 GeV.

An alternative method for breaking the degeneracy between the sign of non-standard interactions and the mass hierarchy is using future medium-baseline reactor experiments like JUNO [18]. While in long baseline and atmospheric neutrino experiments the sensitivity to both NSI and mass hierarchy comes from the matter effect, long baseline reactor experiments can provide an independent measurement of the hierarchy which relies on the interference between the solar and atmospheric neutrino mass scales, with matter effects being unimportant. It is thus in principle possible to gain information about both the mass hierarchy and the sign of $\epsilon_{\mu\tau}$, assuming this is present with the appropriate size.

## VI. CONCLUSION

In this paper we have shown that non-standard neutrino interactions in the mu-tau sector have significant effects on muon neutrino survival probabilities and number of events expected in large atmospheric neutrino detectors. These effects are sign asymmetric and we have derived an analytic expression that can predict points of maximum asymmetry in $(E, \theta)$ parameter space. The increased sensitivity to $\epsilon_{\mu\tau}$ is due to matter interactions, so the largest impact is observed for neutrinos that go through the core of the Earth and this is an indication of where to look for evidences for NSI in experimental data. We have investigated the effects of this sign asymmetry on the mass hierarchy determination. We showed that, even for very small values of $\epsilon_{\mu\tau}$, depending on its sign, it is possible to mimic the wrong mass hierarchy. This emphasizes the importance of considering the framework in which the data is analyzed and the value of consistency checks. We have also discussed how it might be possible to break the degeneracy between the two effects by comparing different energy ranges and possibly by combining the matter-effect based measurements of the hierarchy with those obtained from medium baseline reactor experiments.

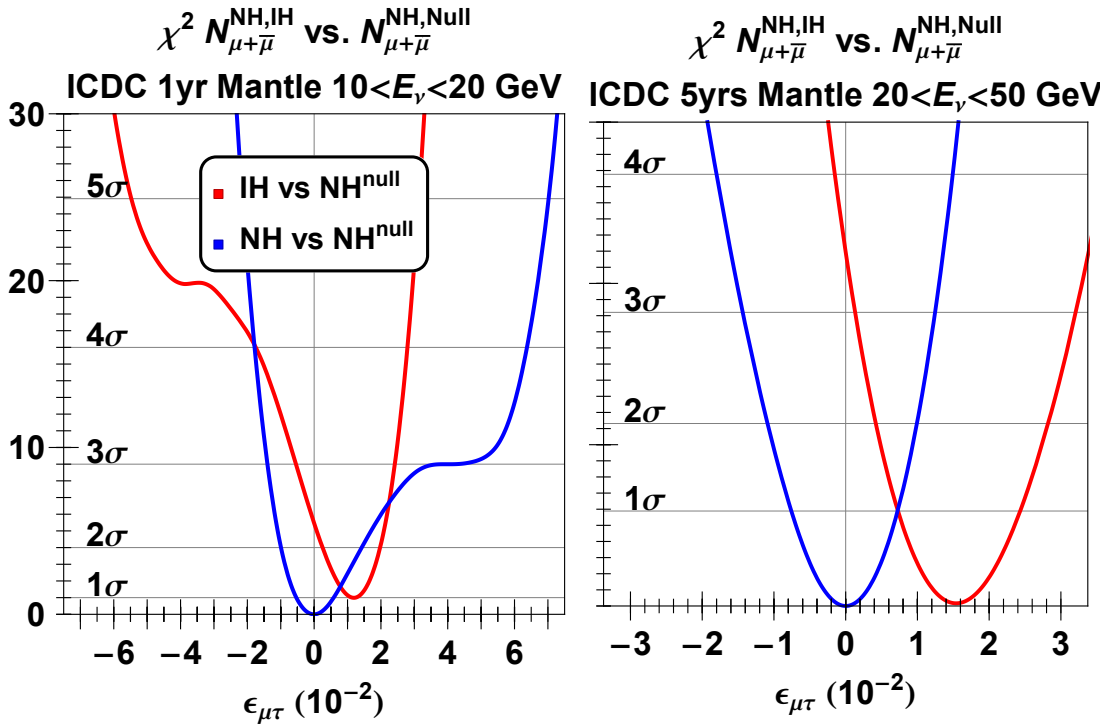

FIG. 10: $\chi^2$ analysis for NH true and for muons traversing the Mantle.

## Acknowledgments

This work was supported in part by the US Department of Energy under contract DE-SC0010534 and by the National Science Foundation under Grant No. NSF PHY11-25915.

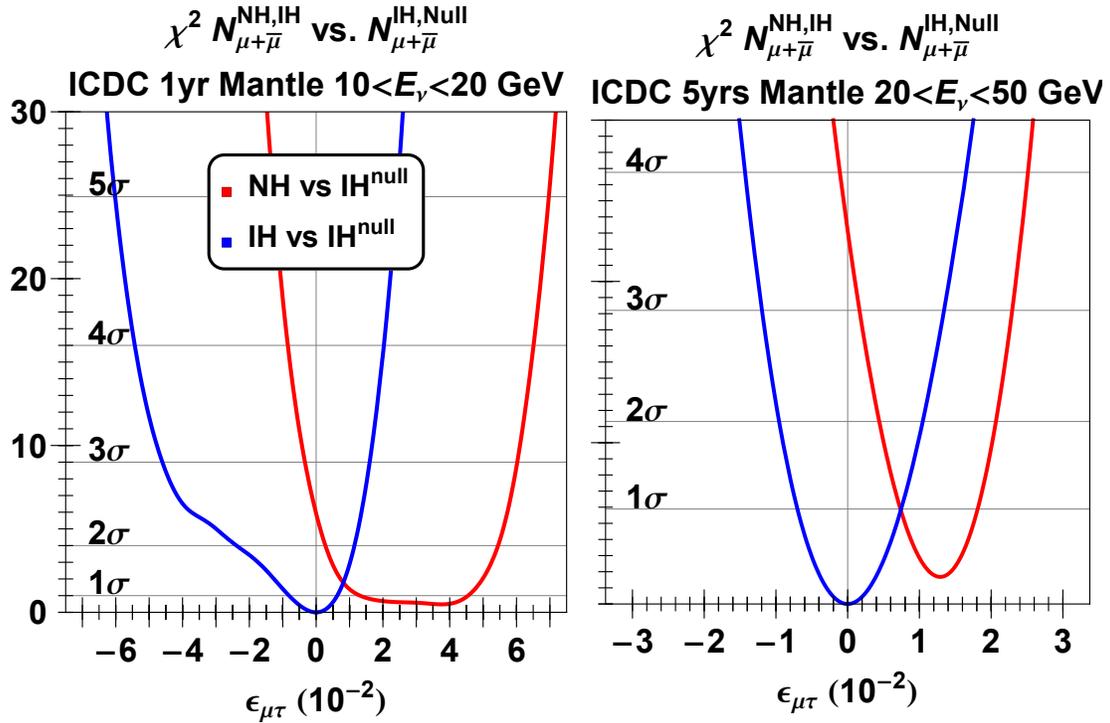

FIG. 11: $\chi^2$ analysis for IH true and for muons traversing the Mantle.